# On the Optimal Scheduling in Pull-based Real-Time P2P Streaming Systems: Layered and Non-Layered Streaming

Abbas Bradai, Toufik Ahmed
CNRS-LaBRI University of Bordeaux-1
351, Cours de la libération
Talence, 33405
{bradai, tad} @labri.fr

*Abstract*—**During the last decade, we witnessed a rapid growth in deployment of pull-based P2P streaming applications. In these applications, each node selects some other nodes as its neighbors and requests streaming data from them. This scheme allows eliminating data redundancy and recovering from data loss, but it pushes the complexity to the receiver node side. In this paper, we theoretically study the scheduling problem in Pull-based P2P video streaming and we model it as an assignment problem. Then, we propose AsSched, new scheduling algorithm for layered streaming, in order to optimize the throughput and the delivery ratio of the system. In second time, we derive an optimal algorithm (NAsSched) for non layered streaming.**
**The results of simulations show that our algorithms significantly outperform classic scheduling strategies especially in stern bandwidth constraints.**

## I. INTRODUCTION

Peer-to-Peer (P2P) architecture is considered as an attractive and scalable solution for video streaming. It does not require internet infrastructure changes and it helps eliminating bandwidth bottleneck at the content source. Nevertheless, P2P systems, especially for real time video streaming (live and video-on-demand), cope with many challenging issues such as overlay construction [1], content retrieval mechanisms (scheduling) [2], and content adaptation [3].

In P2P video streaming systems, the content retrieval mechanism allows a user to receive streaming data blocks (chunks) from other nodes using the constructed overlay. This mechanism plays a leading role in the video streaming process and its efficiency influences the global performance. Two main approaches have been proposed: the pull and the push mechanisms. The pull mechanism is based on the chunks availability at peers: what chunks are available from which neighbor? Thus, a receiver node has to locate the missing chunks and to request them from the appropriate nodes. On the other hand, in the push mechanism, it is the sender nodes which crowd the chunks to the receiver node without any action from this later.
The pull mechanism is considered as very simple and suitable approach as it allows the receiver to cope with two main challenges: eliminating chunks redundancy and recovering from chunks loss. However, it adds complexity to the receiver side because it is responsible for selecting the appropriate chunks to be selected from the appropriate neighbor.

In the context of multi-source overlay network named mesh network, the overlay construction strategy satisfies some quality requirements, such us minimizing delay, maximizing throughput or resiliency, but does not impose how to use the overlay [4]. For example, when maximizing throughput, a node may have parents with high available upload bandwidth, but if they don't have enough or suitable content to send (chunks having past playback deadline for e.g.), the observed throughput will be lower than expected, or useless data will be received by the receiver node. Moreover, when a receiver node requests most parts of chunks from only one of its parents, it will be sensitive to departure or failure of this parent and then may experience significant quality degradation. Hence the system has a weak resiliency. Once the overlay is built, the next step is the scheduling. In this step, the receiver node exchanges information about the available chunks with its neighbors and assigns the task of providing each chunk to a neighbor node. To be efficient, the scheduling has to make the best use of the available bandwidth taking into consideration the availability of chunks in the neighborhood.

The scheduling task is complicated in the context of video streaming since chunks received after their playback deadline are not played and considered as useless chunks. Moreover, in the context of layered video, the task becomes more and more complicated, since an additional constraint should be taken into consideration, namely the layers' dependency. Hence, in layered video coding, video is encoded into a base layer and several enhancement layers, where a higher layer can be decoded only if all related lower layers are available. This is what we call the layers' dependency.

In the literature, the most of related research work tackles the overlay construction problem to improve its efficiency and robustness [5, 6, 7]. But although several schemes were proposed to address the scheduling problem in the context of pull-based architectures. Most of these works are empirical studies or based on queues theories. Indeed, the scheduling strategies adopted in most of the pioneering works mainly include pure random strategy [8], Local Rarest First (LRF) strategy [9] and Round Robin strategy [10] or the queue theory [11]. Furthermore, a few theoretical studies in the literature tackle the optimal stream scheduling. In [12], authors deal with the scheduling problem in data-driven streaming system. They model it as a complicated min-cost network flow problem, and propose a distributed heuristic algorithm to optimize the overall system throughput. In [13], authors propose a 3-stages scheduling approach to request missed chunks, in case of layered streaming.

In this paper we present a new analytical model and its corresponding algorithms to deal with the chunks scheduling problem in Pull-based P2P video streaming, both in case of layered and non-layered video streaming. First, we propose a chunks prioritization strategy in order to represent the urgency of chunks and its layers dependency. Then, we model the problem as an assignment problem and we propose new algorithms to resolve it in order to fully take advantage of bandwidth capacity of the network and to meet the availability of chunks in neighborhood. The rest of this paper is organized as follows: section II formulates the scheduling problem in P2P video streaming, section III models and presents the solution that we propose, section IV presents and discusses the performance evaluation results, and finally, section IV concludes the paper.

## II. CHUNKS SCHEDULING: PROBLEM STATEMENT AND FORMULATION

The basic idea in Pull-based P2P Video Streaming is that the overlay is constructed in such a way to optimize some parameters such as the delay, the bandwidth, etc. Each node in the overlay is connected to a set of neighbors but it is up to the receiver node to ask the chunks from its neighbors. In this paper we assume that the chunks are organized into a sliding window (Figure 1) where chunks beyond the playhead position form the *exchanging window*. Only these chunks are requested if they are not yet received. The missed chunks before playhead position will be no more requested while the chunks received after their playback deadline are not played and considered as useless. Every node periodically sends to all its neighbors a bit vector called *buffer map* (Figure 2), in which each bit represents the availability of a chunk in the sliding window, to announce chunks that it holds. Each node periodically sends requests to its neighbors for the missed chunks in its exchanging window. When a chunk is not received after its request is issued and is still in the exchanging window, it should be requested in the following request period again.

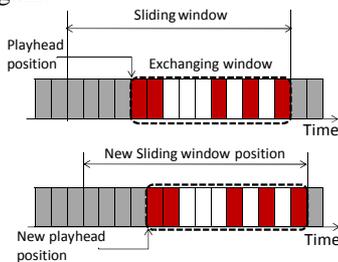
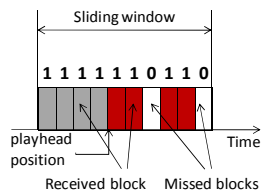

Figure 1: Sliding window mechanism    Figure 2: Buffer map

In order to maximize the throughput of the system, our approach aims to fully take advantage of the receiver nodes' download bandwidth by maximizing the number of chunks that are requested within each scheduling period. Figure 3 illustrates an example of the optimal scheduling problem (in terms of bandwidth utilization). Node 1 is the receiver node. It tries to request missed chunks from its neighbors: nodes 2, 3, 4. Each neighbor expresses the chunks that it holds via a buffer-map. The numbers on the arcs denote the amount of bandwidth that the neighbor node is willing to provide to the receiver node (Node 1) in terms of chunks per unit time. An optimal scheduling schema of this example is represented in Figure 4 where rows represent the nodes and the columns represent the chunks numbers. Chunk 1 is requested from node 4, chunks 2 and 3 from node 2, while chunks 4 and 5 are requested from node 3. This strategy takes full advantage of the available bandwidth of the network. In Figure 5, we represent the result of Round Robin scheduling strategy, described in [14], and applied to the same example. On the contrary of the optimal scheduling strategy, only 4 chunks from 5 can be requested in one unit time in the case of the Round Robin strategy.

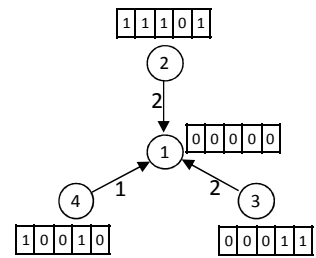

Figure 3: Example of the optimal chunk scheduling problem

| Chunks / Nodes | 1 | 2 | 3 | 4 | 5 |   | Chunks / Nodes | 1 | 2 | 3 | 4 | 5 |
|---|---|---|---|---|---|---|---|---|---|---|---|---|
| 2 | 1 | 1 | 1 | 0 | 1 |   | 2 | 1 | 1 | 1 | 0 | 1 |
| 3 | 0 | 0 | 0 | 1 | 1 |   | 3 | 0 | 0 | 0 | 1 | 1 |
| 4 | 1 | 0 | 0 | 1 | 0 |   | 4 | 1 | 0 | 0 | 1 | 0 |

Figure 4: Optimal Chunk scheduling example     Figure 5: Round robin scheduling example

In addition to fully take advantage of the network capacity, our goal is to ask, in each request period, for chunks having nearest playback deadline first, while taking into consideration the layers dependency of these chunks. The basic idea of our approach is to define priority for each chunk related to its playback deadline and its layers' dependency (in case of layered streaming) and ask in each period for the most priority chunks first, while fully exploiting the receiver node download capacity.

In Table 1 we summarize notations used in the rest of this paper.

| Notation | Description |
|---|---|
| $N$ | The set of receivers' nodes in the overlay |
| $NBR(i)$ | The set of all neighbors of node $i$ |
| $M(i)$ | The set of missed chunks in node $i$ |
| $C_i$ | The current clock on node $i$ |
| $D_j^i$ | The playback time of chunk $j$ on node $i$ |
| $P_{ij}$ | The priority of the chunk $j$ when it is requested from the node $k$. $j \in M(i)$ |
| $R_{ij}^k$ | Boolean variable. $R_{ij}^k = 1$ in case of the node $i$ requests the chunk $j$ from the node $k$, $R_{ij}^k = 0$ otherwise. |
| $L$ | The maximum layer number supported by a node |
| $B_i$ | Vector of download bandwidths between a node $i$ and all its neighbors, $B_i = (b_1, b_2, ..., b_n)$ |
| $r$ | Layers blocks' rate vector, $r = (r_1, r_2, ..., r_n)$ |
| $b_{i,j}$ | Integer, represents the download bandwidth between node $i$ and $j$ (Chunks/time unit) in case of equal size blocks |

Table 1: Notations

### III. MODEL AND SOLUTION

a) Model

The main goal of our scheduling approach is to optimally request the missed chunks in the sliding window of a receiver node from its neighbors. The optimality concerns the request of the higher priority chunks first then the less priority chunks while fully taking advantage of the network capacity. Since in the P2P streaming systems the chunks received after their playback deadline are useless and not played (by consequence the quality of the stream degrades), the priority of the chunk should be closely related to this factor. Initially we consider the *emergency priority* (*EP*) of a chunk defined as its playback deadline. Hence, a chunk with near playback deadline is higher priority than a chunk with far playback deadline. Intuitively because the chunk that is in danger of being delayed beyond the deadline should be more priority than the one just entering the sliding window. The layers dependency is another crucial parameters to be taken into consideration when requesting chunks from neighbors. Indeed, a high layer chunk received without its related lower layers chunks will not be played and considered as useless. Thus, each chunk has an additional key priority, namely the *layer priority* (*LP*).

Therefore, we define chunk $j$'s priority function for the multi-layer scenario as follows:

$$P_{ij} = EP(C_i - D_j^i) + \theta LP(l_j) \quad (1)$$

Where *EP* denotes the emergency priority function related to the remaining time of chunk $j \in M(i)$ till its playback deadline $(C_i - D_j^i)$. The function *LP* represents the layer priority of the layer $l_j$ chunk and the factor $\theta$ is a parameter that can be adjusted for different layers prioritization strategies. Indeed, setting $\theta$ to very low value leads to the prioritization schema represented in Figure 6(a), or the *conservative chunk scheduling*, where the receiver requests always chunks of lower layers first. On the contrary, setting $\theta$ to enough large value, leads to the *aggressive chunk scheduling* schema represented in Figure 6(b). This scheme requests chunks of all layers with lowest time stamp preemptively. While adjusting the value of $\theta$ to a proper value leads to the *zigzag chunk scheduling* in Figure 6(c), which is a trade-off between the two extreme previous schemes.

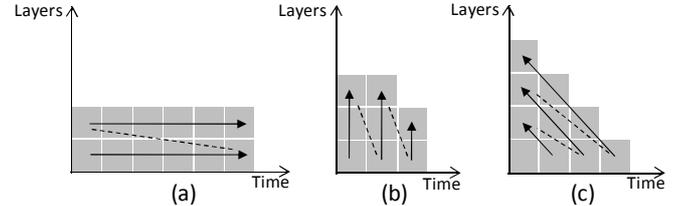

Figure 6: Scheduling strategies in case of layered streaming

We define the Boolean variable $R_{ij}^k$ to denote whether the node $i$ requests the chunk $j$ from the neighbor $k$:

$$R_{ij}^k = \begin{cases} 1, & \text{node } i \text{ should request chunk } j \text{ from neighbor } k \\ 0, & \text{otherwise} \end{cases}$$

Request the most priority chunks first in each request period can be seen as the maximization of the total priority of requested chunks within each request period, i.e.

$$\max \left( \sum_{j \in M(i)} \sum_{k \in NBR(i)} P_{ij} R_{ij}^k \right) \quad (i \in N) \quad (2)$$

Subject to:

$$\sum_{j \in M(i)} R_{ij}^k \leq C_i^k, \quad \sum_{k \in NBR(i)} R_{ij}^k \leq 1 \ (j \in M(i)) \quad (3, 4)$$

Where:

$C_i^k$: Download capacity of the link between the receiver node $i$ and its neighbor $k$.

$M(i)$: Missed chunks in node $i$.

Constraint (3) ensures that the links capacity is not violated, while constraint (4) ensures that a chunk $j$ will be requested from at most one neighbor and no duplicated chunk will be requested to the same neighbor node.

b) Solution

The problem as presented in the previous section can be naturally transformed into an Assignment Problem (AP) [15] where a set of missed chunks $b \in M(i)$ in node $i$ are to be assigned to a set of its neighbors $NBR(i)$ while maximizing the priority sum of the chunks with respect to the download capacity between the receiver node and each of its neighbors. The set of chunks refers to a set of tasks which should be assigned to a set of agents (neighbors nodes) while optimizing the overall cost, which refer to the priority sum of the chunks. In its original version, the AP involves assigning each task to a different agent, with each agent being assigned at most one task, i.e. one-to-one assignment. The other category of the model does assign multiple tasks to the same agent, i.e. one-to-many assignment. In our case, we want to assign one or more chunks to each neighbor, this is why the scheduling in layered streaming matches with the second category of assignment problem, more specifically with the Generalized Assignment Problem (GAP) [15]. This model assumes that each task will be assigned to one agent, but it allows for the possibility that an agent may be assigned more than one task, while recognizing how much of an agent's capacity to do those tasks. Thus, the scheduling problem in layered streaming can be modeled as a GAP and the scheduling of *m* chunks to *n* nodes ($m \geq n$) can be represented by the assignment matrix in Figure 7.

| Chunk / Node | 1 | 2 | … | m-1 | m |
|---|---|---|---|---|---|
| 1 | $P_{i1}$ | $P_{i2}$ | … | $P_{i(m-1)}$ | $P_{im}$ |
| 2 | $P_{i1}$ | $P_{i2}$ | … | $P_{i(m-1)}$ | $P_{im}$ |
| … | … | … | … | … | … |
| n-1 | $P_{i1}$ | $P_{i2}$ | … | $P_{i(m-1)}$ | $P_{im}$ |
| n | $P_{i1}$ | $P_{i2}$ | … | $P_{i(m-1)}$ | $P_{im}$ |

(Nodes' reliability)

Figure 7: assignment matrix -GAP

The GAP is known to be NP-hard problem. In the following section we propose a new heuristic (AsSched) to resolve it and perform the chunk scheduling in Pull-based P2P streaming architectures.

*Algorithm*

In order to construct a solution for the scheduling problem in layered video, modeled as GAP, we consider an algorithm *A* for the knapsack problem (Let be the Harmony-search algorithm[16]). First, we reorganize the rows of the assignment matrix based on neighbors' reliability (Figure **7**) in order to assign chunks to the higher reliable nodes first then the lower reliable ones.

Since our algorithm modifies the assignment matrix, we use the notation $M_j$ to note the assignment matrix at the $j^{th}$ recursive call (*j* initialized to 0) of the following *LineProcessing* (*j*) procedure:

1. Run the Algorithm *A* on the row *j* with respect to the download bandwidth of the node *i* ($b_i$) and chunks' size *r*, and let $S_j$ be the set of selected chunks returned.
2. Set all the priorities of the chunks corresponding to the selected chunks to -*M* (*M* is a large positive number), i.e. $\forall k \in S_j, \forall x : M_i(x,k) = -M$
3. If *j*<*n* (with *n=card* (*NBR(i)*) )
   - Remove the row of the node *j* from $M_j$ and set $M_{j+1}=M_j$
   - Perform *LineProcessing* (*j+1*) and let $S_{j+1}$ be the returned chunks list, and let $\overline{S} = S_j \cup S_{j+1}$
   - Return $\overline{S}$
   Else return $\overline{S}$

The solution, for the layered video, proposed in this section, can be easily extended to the non-layered video, by considering the number of layers equal to one and setting the dependency priority function *EP* = 0. But, can we do better?

c) Special case: Non-layered streaming

In this section we propose to adapt and to simplify the solution presented in the last section to the non-layered video streaming. Initially, the priority function $P_{ij}$ is simplified to the emergency priority *EP*. In addition, we assume that the non-layered video is subdivided into chunks of equal size. It is hard to consider this assumption in the case of layered video, especially in the case of SVC [17] where the video stream is subdivided into NALs (Network Abstraction Layer) of different sizes. Consequently, the scheduling problem in layered video streaming can be modeled as one-to-one assignment problem, more especially as *m*-cardinality assignment problem [15], defined as the assignment of *m* jobs among *n* to *m* agents. To do that, each neighbor node is represented in the assignment matrix of the receiver node *i* by $b_{i,j}$ rows, i.e. for each node corresponds $b_{i,j}$ virtual nodes, each one with a capacity of one chunk per time unit (Figure 8).

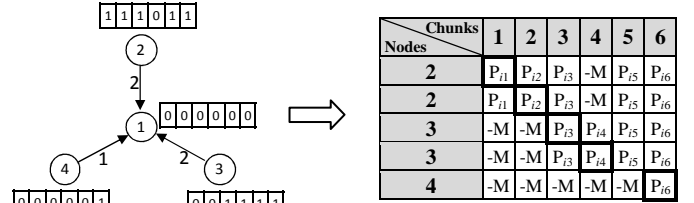

Figure 8: m-cardinality assignment matrix example

In order to resolve this problem we propose to transform it, first, to a one-to-one classic assignment problem (square matrix: Figure 10), and then apply the Hungarian algorithm [18] to get the optimal scheduling. The Hungarian algorithm is a powerful combinatorial optimization algorithm, which solves a classical AP in polynomial time. It is applicable, exclusively, to square assignment matrix.

| Chunks / Nodes | 1 | 2 | … | *l* |
|---|---|---|---|---|
| 1 | $P_{i1}$ | $P_{i2}$ | … | $P_{im}$ |
| 2 | $P_{i1}$ | $P_{i2}$ | … | $P_{im}$ |
| … | … | … | … | … |
| n | $P_{i1}$ | $P_{i2}$ | … | $P_{im}$ |

Figure 9: m-cardinality assignment matrix

| Chunks / "Nodes" | 1 | 2 | … | *l* |
|---|---|---|---|---|
| 1 | $P_{i1}$ | $P_{i2}$ | … | $P_{il}$ |
| 2 | $P_{i1}$ | $P_{i2}$ | … | $P_{il}$ |
| … | … | … | … | … |
| n | $P_{i1}$ | $P_{i2}$ | … | $P_{il}$ |
| n+1 | L | L | L | L |
| … | L | L | L | L |
| *l* | L | L | L | L |

Figure 10: Transformed m-cardinality assignement matrix

**Transformation rules**

The following steps are performed to build the new square assignment matrix (Figure 10) of a receiver node *i*:

a) For each nodes $j \in NBR(i)$ add $b_{i,j}$ rows to the matrix, (matrix of *m* rows, where $m = \sum_{j \in NBR(i)} b_{i,j}$ )
b) For each missed chunk in node *i* add a column to the matrix (matrix of *l* columns)
c) The value of *Cell*(*k*, *j*) is the chunk *i*'s priority, i.e. *Cell*(*k*,*j*) =$P_{ij}$, if the node *k* holds the chunk *j*, -*M* otherwise (*M* is a big positive number).
d) If the matrix is not square, i.e. *l* > *m*: append *x* = *l-m* virtual nodes to the assignment matrix. Set the *Cell*(*k*, *j*) value to L for each row *k*∈ {*l-m*+1, *l-1*}, where *L* is a positive number (Figure 10).

After applying these rules, we transform the formulation (2) into its corresponding assignment problem represented by a square matrix ($l \times l$) composed of $l$ chunks to be assigned to $l$ "nodes". Hence, the Hungarian algorithm can be applied to get the optimal chunk scheduling.

Formally, the assignment problem (2) can be rewritten into the following assignment problem:

$$Max\left(\sum_{k=1}^{k=m}\sum_{j=1}^{j=l} P_{ij}^k R_{ij}^k + \sum_{k=m+1}^{k=l}\sum_{j=1}^{j=l} LR_{ij}^k\right) \quad (i \in N) \quad (5)$$

Subject to

$$\sum_{k=1}^{k=l} R_{ij}^k = 1 \quad (1 \le j \le l), \quad \sum_{j=1}^{j=l} R_{ij}^k = 1 \quad (1 \le i \le l) \quad (6)$$

**Theorem**

Let $\overline{R}_{ij}^k (1 \le j,k \le l)$ be an optimal solution to the assignment problem (5), then $\overline{R}_{ij}^k$ is an optimal solution to the $m$-cardinality assignment problem (2).

**Proof.**

Suppose that $\overline{R}_{ij}^k (1 \le j,k \le l)$ is not an optimal solution to the $m$-assignment problem (2), then there exists a feasible solution $\hat{R}_{ij}^k (1 \le j \le l, \ 1 \le k \le m)$ of (2) which verifies:

$$Max\left(\sum_{k=1}^{k=m}\sum_{j=1}^{j=l} P_{ij}\hat{R}_{ij}^k\right) > Max\left(\sum_{k=1}^{k=m}\sum_{j=1}^{j=l} P_{ij}^k \overline{R}_{ij}^k\right)$$

Without loss of generality, we assume that

$$\sum_{k=1}^{k=m}\sum_{j=1}^{j=m} \hat{R}_{ij}^k = m$$

And

$$\hat{R}_{(m+1)(m+1)}^k = \hat{R}_{(m+2)(m+2)}^k = ... = \hat{R}_{ll}^k = 1$$

Since the cost $P_{ij} = L$ for $m+1 \le k \le l$ and $1 \le j \le l$, we have:

$$Max\left(\sum_{k=1}^{k=m}\sum_{j=1}^{j=l} P_j^i \hat{R}_{ij}^k + (l-m)L\right) > Max\left(\sum_{k=1}^{k=m}\sum_{j=1}^{j=l} P_{ij}^k \overline{R}_{ij}^k + (l-m)L\right)$$

Which is in contradiction with that $\overline{R}_{ij}^k (1 \le i,j \le l)$ is an optimal solution to the assignment problem (5).

IV. PERFORMANCE EVALUATION

As abovementioned, there are three main steps for building streaming applications in overlay networks. In this paper we focus on the streaming scheduling step. For that reason we use in all our simulation a simple algorithm for overlay construction: each node randomly selects its neighbors so that a random graph is constructed. The overlay is composed of 500 nodes and each node has 15 neighbors. Each node estimates the bandwidth allocated from a neighbor with the traffic received from it in previous 5 periods using Adaptive Linear Prediction method [19]. We have performed extensive simulations using Simulink-Matlab simulations [20].

The performance of our algorithm is compared to the performance of the three scheduling methods described earlier in section II, namely Random strategy (RND), Local Rarest First (LRF) and Round Robin (RR). We consider three categories of peers: 40% users with 512Kbps, 30% with 1Mbps and 30% with 2Mbps, and for all users, the upload bandwidth capacity is half of the download bandwidth.

To evaluate the performance under multilayer scenario, we define the delivery ratio at layer $l$ as the average delivery ratio at layer $l$ among all the nodes that can play layer $l$. A chunk of layer $l$ is considered as well received if and only if all its related chunks of lowers layers to $l$ are already received no later than the playback deadline. We set the emergency priority defined in (1) as $EP_{ij}(C_i - D_j^i) = 10^{(C_i - D_j^i)}$ and we set the layer priority as $P_L(l_j) = 10^{(L-l_j)}$ to ensure that the lower layers have much larger priority than the upper layers. For the four methods, we adopt the conservative approach described in section III. This is why we set the parameter $\theta$ to a very low value $\theta = 10^{-L}$.

We first encode the video into 12 layers and set the rate of each layer at 100 Kbps. Figure 11 describes the delivery ratio at each layer. We note that AsSched is fairly good. In lower layers, most of the delivery ratio is nearly 1 and most in higher layers is also above 0.9. The RR has much more better delivery ratio at lower layers than higher layers. But, the delivery ratio at all layers is not so good as the proposed algorithm. We note that the LRF strategy has even higher delivery ratio than the RR strategy. Finally, the random strategy has the poorest performance. As shown in Figure 11, our algorithm outperforms other strategies with a gain of 10%-50% in most layers.

In order to show the importance of different layers encoding schemes, we encode the video into 6 layers. In Figure 12, we note that the delivery ratio of each layer is nearly similar to that in 12 layers encoding scenario. AsSched is still the best among all the three others methods. However, we note that the delivery ratio of all the methods is little higher than in the case of 12 layers. This is due to the fact that encoding the video into six layers allows nodes to allocate all their bandwidth to lower layers, however in the second case, some bandwidth will be dedicated to the higher layers (higher than 6).

To evaluate the performances under single layer scenario, we define the *delivery ratio* to represent the number of chunks that arrive at each node before their playback deadline over the total number of chunks encoded. The average delivery ratio represents the throughput of the whole system and reflects the average quality observed by users.

In Figure 13 we study the performance of NAsSched compared to RR, LRF and RND under different streaming rate conditions. We set the exchanging window size to 10 seconds and the video chunks have the same size of 10 Kbits.

We note that when the streaming rate is low (250Kbps for e.g.), all the algorithms have high delivery ratio. We explain this by the fact that a stream chunk has more chance to be re-scheduled before the playback deadline in the case of low streaming rate. However, when the streaming rate increases, the performance of the three compared algorithms decreases fast. At the rate of 500Kbps, our algorithm has a delivery ratio of 95% which outperforms the other three methods by gains of about 15% and 60%.

In Figure 14 we set the exchanging window size to 3 seconds. We note that with a smaller window size, the delivery ratio of all algorithms decreases. Since the request period is set to 1 second, most of the chunks can only be requested repeatedly for 3 times. Our proposed algorithm outperforms the others because it fully takes advantage of the network capacity and chooses the most appropriate neighbors to ask from in each period.

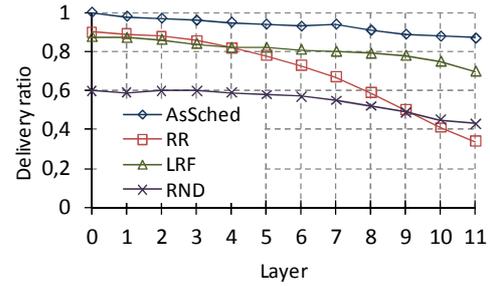

Figure 11: Multi-layer scheduling - 12 layers

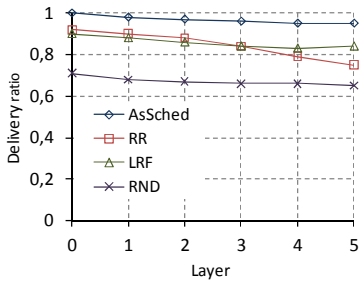

Figure 12: Multi-layer scheduling - 6 layers

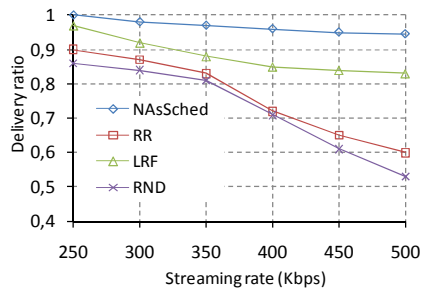

Figure 13: Single layer scheduling - exchanging windows of 10 seconds

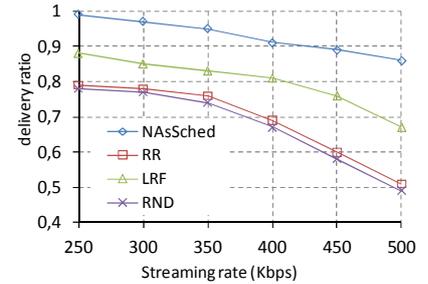

Figure 14: Single layer scheduling - exchanging windows of 3 seconds

## V. CONCLUSION

In this paper, we tackle the optimal scheduling problem in pull-based real-time streaming systems in multilayer streaming scenarios. We model the problem as a Generalized Assignment Problem and we propose a heuristic to resolve it. Then, in second time, we adapt the solution to non-layered streaming and we model it as *m*-cardinality assignment problem and we propose a new solution for this problem. The simulation results show that the proposed solutions outperform the traditional strategies by about 15 to 60 percent both in single and multilayer streaming.